\documentclass[aps,superscriptaddress,showpacs,amsmath,amssymb,longbibliography,twocolumn, floatfix,prd]{revtex4-2}
\usepackage[T1]{fontenc} 
\usepackage{color} 
\usepackage[english]{babel} 
\usepackage{verbatim} 
\usepackage{float} 
\usepackage{graphicx} 
\usepackage[pdftex, colorlinks=true, linkcolor=blue, citecolor=blue,urlcolor=blue]{hyperref}
\usepackage{multirow}
\usepackage{array}
\usepackage[utf8]{inputenc}
\usepackage{natbib}
\usepackage[dvipsnames]{xcolor}

\usepackage[normalem]{ulem}

\newcolumntype{P}[1]{>{\centering\arraybackslash}p{#1}}

\newcommand{\MT}[1]{{\color{black}#1}} 
\newcommand{\FP}[1]{{\color{black}#1}} 

 \pdfpageheight\paperheight \pdfpagewidth\paperwidth

\usepackage{pdfcomment} 
\usepackage{environ} 
\RenewEnviron{comment}{\pdfcomment{\BODY}}

\makeatother

\begin{document}

\title{Phonon-assisted carrier cooling in $h$-BN/graphene van der Waals heterostructures}

\author{Sangkha Borah}
\affiliation{Okinawa Institute of Science and Technology Graduate University, Onna-son, Okinawa 904-0495, Japan}
\author{Dinesh Yadav}
\affiliation{Institute of Physics, University of Augsburg, 86135 Augsburg, Germany}
\affiliation{Okinawa Institute of Science and Technology Graduate University, Onna-son, Okinawa 904-0495, Japan}
\author{Maxim Trushin}
\affiliation{Institute for Functional Intelligent Materials, National University of Singapore, 117544, Singapore}
\affiliation{Department of Materials Science and Engineering, National University of Singapore, 117575 Singapore}
\affiliation{Centre for Advanced 2D Materials, National University of Singapore, Singapore 117546}
\author{Fabian Pauly}
\email{fabian.pauly@uni-a.de}
\affiliation{Institute of Physics, University of Augsburg, 86135 Augsburg, Germany}
\affiliation{Okinawa Institute of Science and Technology Graduate University, Onna-son, Okinawa 904-0495, Japan}

\date{\today} 

\begin{abstract}
Being used in optoelectronic devices as ultra-thin conductor-insulator junctions, detailed investigations are needed about how exactly $h$-BN and graphene hybridize. Here, we present a comprehensive {\em ab initio} study of hot carrier dynamics governed by electron-phonon scattering at the $h$-BN/graphene interface, using graphite (bulk), monolayer and bilayer graphene as benchmark materials. In contrast to monolayer graphene, all multilayer structures possess low-energy optical phonon modes that facilitate carrier thermalization. We find that the $h$-BN/graphene interface represents an exception with a comparatively weak coupling between low-energy optical phonons and electrons.
As a consequence, the thermalization bottleneck effect, known from graphene, survives hybridization with $h$-BN, but is substantially reduced in all other bilayer and multilayer cases considered. In addition, we show that the quantum confinement in bilayer graphene does not have a significant influence on the thermalization time compared to graphite and that bilayer graphene can hence serve as a minimal model for the bulk counterpart.

\end{abstract}
\keywords{Graphene, $h$-BN, interface, van der Waals heterostructures, hot carriers, thermalization, density functional theory, electron-phonon interaction, optoelectronics}

\maketitle

\section{Introduction}
How does an excited electron or hole lose its energy? This question is at the heart of photovoltaic technology, as the answer offers a hint for assessing the prospects of a given material in solar cell applications~\cite{Das2019}. Van der Waals (vdW) heterostructures~\cite{Geim2013} of atomically thin crystals emerge as a flexible platform that allows for adjustment of the scattering channels and subsequent control of optoelectronic~\cite{tian2016optoelectronic,ponraj2016photonics,Castellanos-Gomez2016,Wang2012} and photovoltaic~\cite{Das2019,Wang2018,Bati2019,Cheng2018} properties. The simplest {realizations} may be bilayer heterostructures made of a two-dimensional (2D) semiconductor and graphene~\cite{graphene-MoS2, WSe2-graphene, graphene-MoS2-older}. The honeycomb symmetry of graphene's carbon lattice results in a linear electron dispersion and high optical phonon excitation energies, which all together leads to a slow transfer of photocarrier energy to the lattice~\cite{Konig-Otto2016,Mihnev2016}. On the semiconducting side, however, the electronic dispersion at the band edge is parabolic, and electron-phonon (EP) scattering typically turns out to be much stronger than in graphene \cite{abrikosov2017fundamentals,Song2015,Brida2013,Konig-Otto2016}. This can lead to the situation that photoexcited carriers in the semiconducting layer dissipate their energy much faster than in the attached graphene. Such an asymmetric behavior of photocarrier energy dissipation on the two sides of a heterojunction can create a temperature gradient, resulting in an intrinsic photothermoelectric effect~\cite{graphene-MoS2}. Combining different materials in the prototypical bilayer stacks is expected to reveal the distinct properties of such heterojunctons.  Along these lines, a recent review~\cite{graphene-semicond-review} suggests six strategies for improving the performance of vdW optoelectronic devices. To make a deliberate choice between the possible strategies, we need to obtain a better understanding of the photocarrier evolution in vdW heterostructures, which may strongly differ from that in isolated 2D layers.

A common simplifying assumption in modeling optoelectronic properties of vdW bilayers is that the two 2D materials are in electrical contact, but that their single-layer properties remain otherwise unaffected~\cite{MISdiode}. Following the electron affinity rule~\cite{nelson2003physics} (also known as the Schottky-Mott rule for metal-semiconductor junctions~\cite{PhysRev.71.717}), one can figure out the energy band alignment across the interface, see Refs.~\cite{trushin2018theory,cao2019janus,cao2020electrical,cao2020tunable,cao2021two} for recent bilayer examples. This simple approach might not necessarily provide a complete physical picture, especially for 2D materials, as the interfacial region in vdW bilayers represents an integral part of the structure. Hence, vdW heterostructures made of two monolayers should be seen as a new single material possessing properties qualitatively different from those of each individual monolayer. 

Since photocarrier relaxation processes take place on femtosecond or picosecond time scales, they are  challenging to {probe} experimentally~\cite{Brida2013,Breusing2011}. In addition, atomically thin samples require a high sensitivity of optical detection. Important recent advances on the experimental side have been achieved by breakthroughs in time-resolved spectroscopy, notably time- and angle-resolved photoemission spectroscopy (TR-ARPES) and pump-probe techniques~\cite{Liu2010,Shang2011,PhysRevLett.111.027403,Carrier_dynamics_free_carriers}. On the theoretical side, the study of ultrafast hot carrier relaxation processes requires state-of-the-art techniques~\cite{Park:PRL2009,Bernardi2014}, which solve the many-body physics in the time domain~\cite{Bernardi:EPJB2016}. Recent developments in {\em ab initio} computational modelling techniques are capable of resolving individual scattering processes and provide insights beyond experiments~\cite{Giustino2017, Bernardi2015, Winzer2010, Bernardi:EPJB2016, Yadav2019, PRB2011Malic, Park:PRL2009, Mauri_PhysRevB.76.205411,PhysRevB.85.241404}.

Photocarrier dynamics in graphene has been studied experimentally~\cite{Brida2013,Breusing2011,Shang2011,Winnerl2011,Gierz2017} as well as theoretically using empirical models~\cite{Mihnev2016,Konig-Otto2016,Winnerl2011,Butscher2007} and recently in terms of a parameter-free first principles approach~\cite{Yadav2019,Tong2021}. Graphene is typically characterized on a substrate such as silicon oxide (SiO\textsubscript{2})~\cite{Chen2008,PhysRevLett.102.076102} or silicon carbide (SiC)~\cite{Mendes_de_Sa_2012}. Rough substrate surfaces and impurities on the surface or the graphene itself act as charge traps and lead to an uneven charge distribution that reduces the carrier mobility in graphene~\cite{Martin2008, Chen2008}. On the other hand, being atomically thin, $h$-BN exhibits a smooth surface almost free from dangling bonds and charge traps~\cite{Dean2010}, a high-temperature stability~\cite{melting_hbn}, and a nearly commensurate lattice structure to graphene. These qualities have proven $h$-BN to be an ideal substrate for graphene. Devices based on $h$-BN/graphene have shown high carrier mobilities~\cite{Dean2010}, ballistic transport~\cite{Wantabe1, Wantabe2}, and quantum Hall effects~\cite{Dean2013,Ponomarenko2013}. There are a number of theoretical studies on electronic and phononic properties as well as electron-phonon couplings for the $h$-BN/graphene system~\cite{Jung2015, Giovannetti:PRB2007,gapvsd,Wahib:PRMater2020,Slotman:AnnPhys2014}. The role of optical and acoustical phonons in carrier thermalization is studied well in graphene~\cite{Tong2021,Yadav2019, Winnerl2011}, but the effects of substrate phonons and an increasing number of graphene or insulating $h$-BN layers are yet to be explored. These are aspects that we address in the present work.

Here, we use {\em ab initio} methods to uncover the effects of interlayer hybridization on photocarrier dynamics in vdW heterostructures prototyping graphene and $h$-BN as constituting monolayers. As compared to the monolayer, new interlayer phonon modes arise in bilayer and multilayer structures. Caused by the weak interlayer coupling, low-energy optical modes will emerge in particular from what are acoustical or flexural modes in the decoupled monolayers. The photocarrier thermalization bottleneck in graphene occurs for excitation energies below the high energy optical phonon modes~\cite{Yadav2019}, i.e.\ electronic excitation energies below around 150~meV. It is thus of particular interest to explore, how these new optical phonon modes at low energies impact carrier dynamics and whether they remove the thermalization bottleneck. We will show that photocarrier dynamics nearly preserves monolayer features, if graphene is placed on $h$-BN. In contrast the bottleneck is strongly reduced in bilayer graphene, which resembles graphite from the photocarrier thermalization point of view.

The rest of the paper is organized as follows. We present our theoretical approach to study photocarrier dynamics in Sec.~\ref{sec:theory}. Results are discussed in Sec.~\ref{sec:results}, followed by the summary and outlook in Sec.~\ref{sec:conclusion}.

\section{Theoretical approach}\label{sec:theory}
\subsection{Electronic and phononic properties}
We combine density functional theory (DFT)~\cite{Kohn1965,Hohenberg1964} with density functional perturbation theory (DFPT)~\cite{Baroni2001} to determine the EP scattering. With the Kohn-Sham eigenstates $|n \mathbf{k} \rangle$ and the derivative of the self-consistent potential $\partial_{p\mathbf{q}}V$, EP coupling matrix elements can be computed as,
\begin{equation}
g_{mn,p}(\mathbf{k,q})=\frac{1}{\sqrt{2\omega_{p\mathbf{q}}}}\left\langle m\mathbf{k+q}|\partial_{p\mathbf{q}}V|n\mathbf{k}\right\rangle,\label{eq:EPcouplings}
\end{equation}
where $m, n$ denote electronic band indices, and $\hbar\omega_{p\mathbf{q}}$ is the energy of the phonon mode $p$ at wave vector $\mathbf{q}$~\cite{Bernardi:EPJB2016}. The matrix element $g_{m n,p}(\mathbf{k,q})$ defines an electronic transition from the initial state $|n\mathbf{k}\rangle$ to the final state $|m\mathbf{k+q}\rangle$ by scattering from the phonon $p\mathbf{q}$. We determine the electron self-energy $\Sigma_{n\mathbf{k}}(T)$ due to EP interaction within the Migdal approximation~\cite{Bernardi:EPJB2016,Giustino2017},
\begin{eqnarray}
\nonumber
    \Sigma_{n\mathbf{k}}(T) &= &\sum_{m,p}\int_{\text{BZ}}\frac{d^3q}{\Omega_{\text{BZ}}}\left\vert
    g_{mn,p}(\mathbf{k},\mathbf{q})\right\vert^{2} \\ 
    & \times & \Bigg[ \frac{N_{p\mathbf{q}}(T)+f^{(0)}_{m\mathbf{k+q}}(T)}{\varepsilon_{n\mathbf{k}}-(\varepsilon_{m\mathbf{k+q}}-\varepsilon_{\text{F}})+\hbar\omega_{p\mathbf{q}} +\text{i}\eta}\label{eq:self-energy}\\ 
    & & +\frac{N_{p\mathbf{q}}(T)+1-f^{(0)}_{m\mathbf{k+q}}(T)}{\varepsilon_{n\mathbf{k}}-(\varepsilon_{m\mathbf{k+q}}-\varepsilon_{\text{F}})-\hbar\omega_{p\mathbf{q}}+\text{i}\eta} \nonumber
    \Bigg],
  \end{eqnarray}
where $\varepsilon_{\text{F}} = 0$ is the Fermi energy, $f^{(0)}_{n\mathbf{k}}(T)=1/[\exp(\frac{\varepsilon_{n\textbf{k}}-\varepsilon_{\text{F}}}{k_{\text{B}}T})+1]$
is the Fermi-Dirac distribution,
$N_{p\mathbf{q}}(T)=1/[\exp(\frac{\hbar\omega_{p\mathbf{q}}}{k_{\text{B}}T})-1]$ is the Bose-Einstein distribution, $\Omega_{\text{BZ}}$ is the volume of the Brillouin zone (BZ), and $\eta$ is a small broadening parameter. The first term in the parentheses in Eq.~(\ref{eq:self-energy}) represents the self-energy contribution from the absorption of phonons and the second one stems from their emission. In order to accurately map the EP scattering events in the whole BZ, it is important that the {self-energy} calculations are performed on fine grids over the electronic ($\mathbf{k}$) and phononic  ($\mathbf{q}$) states. \FP{For this purpose we use an interpolation scheme based on maximally localized Wannier functions (MLWF)~\cite{Mostofi2014} in the software  package \textsc{perturbo}~\cite{Zhou2020perturbo}.}

To determine ground state electronic and phononic properties, DFT within the local density approximation (LDA), as implemented in \textsc{quantum espresso}~\cite{Giannozzi2017}, is used. \FP{Our calculations are performed with Slater exchange and Perdew-Wang correlation (exchange-correlation functional "SLA PW NOGX NOGC") \cite{Giannozzi2017}.} \FP{Core electrons are modelled through optimized norm-conserving Vanderbilt pseudopotentials~\cite{Hamann2013}.} We employ plane wave (PW) basis sets with a kinetic energy cutoff of 90~Ry and a charge density cutoff of 360~Ry. \FP{The energy smearing is set to 0.02~Ry.} We optimize unit cells and atoms therein using the Broyden-Fletcher-Goldfarb-Shanno algorithm \cite{Fletcher2000}, applying tight convergence criteria on forces ($10^{-6}$~Ry/a.u.) and  total energies ($10^{-8}$~Ry). \FP{To avoid artificial interactions, we separate periodic images through a vacuum of 16~\textup{\AA} (monolayer graphene) or 20~\textup{\AA} (bilayer graphene and $h$-BN/graphene) along the $z$-direction for the 2D systems and truncate the Coulomb interaction in the out-of-plane direction (flag "{assume\textunderscore{}isolated=2D}" \cite{Sohier2017Aug}).} VdW corrections are included for bilayer systems and graphite according to a scheme of Grimme~\cite{Grimme_vdw,Giannozzi2017,Barone_vdw}. Calculations of electronic wavefunctions in the BZ are performed on a $\Gamma$ centered $36\times 36 \times 1$ coarse $\mathbf{k}$ mesh for the 2D systems, while we choose a $12\times 12\times 12$ $\mathbf{k}$ mesh for bulk graphite. 

We compute EP couplings in Eq.~(\ref{eq:EPcouplings}) and the self-energy in Eq.~(\ref{eq:self-energy}) with the help of the \textsc{perturbo} code~\cite{Zhou2020perturbo}. For that purpose the integration in Eq.~(\ref{eq:self-energy}) is carried out on a $1800 \times 1800 \times 1 $ $\mathbf k$ mesh for the 2D systems, along with a $\mathbf{q}$ mesh containing $10^7$ points, which we sample from a uniform distribution.  \FP{In addition we truncate Coulomb interactions for the 2D systems (flag "system2d=True").} For bulk graphite, we select a $1800 \times 1800 \times 6$ $\mathbf{k}$ mesh with $10^7$ uniformly distributed $\mathbf{q}$ points. Together with $\eta=10$~meV, we find this to be adequate to achieve converged results. Beyond that we assume that EP couplings are sufficiently weak to neglect their influence on electron wavefunctions and phonon dynamical matrices~\cite{Zhou2020perturbo}. Anharmonic effects, causing the renormalization of phononic properties, are also disregarded for simplicity~\cite{Giustino2017}.

\subsection{Time-evolution of excited charge carriers}
We describe the time evolution of the electronic occupation $f_{n\textbf{k}}(t,T)$ in terms of the Boltzmann equation within the relaxation time approximation (RTA)~\cite{Lundstrom2000} via
\begin{equation}
  \frac{df_{n\textbf{k}}(t,T)}{dt}=-\frac{f_{n\textbf{k}}(t,T)-f^{\text{(0)}}_{n\textbf{k}}(T)}{\tau_{n\textbf{k}}{(T)}}\label{eq:Boltzmann}.
\end{equation}
Assuming the excitation to take place at $t=0$, the solution is
\begin{equation}
    f_{n\textbf{k}}(t,T)=f^{(0)}_{n\textbf{k}}(T)+e^{-\frac{t}{\tau_{n\textbf{k}}{(T)}}}[f_{n\textbf{k}}(0,T)-f^{\text{(0)}}_{n\mathbf{k}}(T)],\label{eq:Boltzmann-solution}
\end{equation}
where $f_{n\textbf{k}}(0,T)$ denotes the initial occupation of hot carriers and $f^{\text{(0)}}_{n\mathbf{k}}(T)$ is the previously defined Fermi-Dirac distribution. We determine the temperature-, band- and wavevector-dependent scattering rates $\tau_{n\mathbf{k}}^{-1}{(T)}$ that appear in Eqs.~(\ref{eq:Boltzmann}) and (\ref{eq:Boltzmann-solution}) from the imaginary part of the self-energy in Eq.~(\ref{eq:self-energy}) as
\begin{equation}
    \tau_{n\mathbf{k}}^{-1}(T)=\frac{2}{\hbar} \text{Im}[\Sigma_{n\textbf{k}}(T)]
    \label{eq:tau-nk}.
\end{equation}

In our analysis, the initial distribution {$f_{n\textbf{k}}(0,T)$} of hot carriers is generated using a Fermi-Dirac distribution at temperature $T$ and Gaussian distributions with peak values at excitation energies $\pm (\xi + \Delta/2)$ above and below the Fermi level for holes and electrons, respectively. In detail,
\begin{equation} 
  f_{n\textbf{k}}(0,T)=f^{(0)}_{n\textbf{k}}(T)\begin{cases}+\lambda_{\text{e}}{(\xi)}e^{-\frac{(\varepsilon_{n\textbf{k}}-\xi-\Delta/2)^{2}}{2\sigma^{2}}},
  & \varepsilon_{n\textbf{k}}\geq\varepsilon_{\text{F}},\\ 
  -\lambda_{\text{h}}{(\xi)}e^{-\frac{(\varepsilon_{n\textbf{k}}+\xi+\Delta/2)^{2}}{2\sigma^{2}}},
  & \varepsilon_{n\textbf{k}}<\varepsilon_{\text{F}}.\end{cases}\label{eq:initial-distribution}
\end{equation} 
In the above equation, $\Delta$ is the bandgap of the material, $\sigma=8.47$~meV is a small broadening, {and $\xi$ is the excess energy}. We furthermore define the full population as
\begin{equation}
P(E,t,T)=\sum_{n\mathbf{k}}\delta(E-\varepsilon_{n\mathbf{k}}) \times
\begin{cases}f_{n\mathbf{k}}(t,T),&
E\geq\varepsilon_{\text{F}},\\ [1-f_{n\mathbf{k}}(t,T)],&
E<\varepsilon_{\text{F}},\end{cases} \label{eq:population}
\end{equation}	
the time-independent equilibrium population $P_0(E,T)$ is obtained from Eq.~(\ref{eq:population}) by replacing $f_{n\mathbf{k}}(t,T)$ with $f^{(0)}_{n\mathbf{k}}(T)$, and the difference yields $P_\text{exc}(E,t,T)=P(E,t,T)-P_0(E,T)$. Electron and hole densities are related to the energy-, time- and temperature-dependent photoexcited carrier population $P_\text{exc}(E,t,T)$ at $t=0$ through \begin{equation}
    \begin{split}
    n_\text{e}\alpha&=\int_{\varepsilon_{\text{F}}}^{\varepsilon_\text{max}}P_\text{exc}(E,{0},T) dE,\\
    n_\text{h}\alpha&=\int_{-\varepsilon_\text{max}}^{\varepsilon_{\text{F}}}P_\text{exc}(E,{0},T) dE,
    \end{split}\label{eq:n_eh}
\end{equation} \FP{where $\alpha$ is the in-plane area of the unit cell for monolayer graphene, bilayer graphene and $h$-BN/graphene, or the volume of the unit cell for graphite. Furthermore, in the expressions we integrate from the Fermi energy $\varepsilon_\text{F} = 0$ to a sufficiently large $\varepsilon_\text{max} = 4$~eV (see also Fig.~\ref{fig:bands}). For each material, we choose initial photoexcited carrier densities of electrons and holes to be equal, $n_\text{h} = n_\text{e}$, and constant with regard to different temperatures $T$ and excess energies $\xi$ considered.}
\FP{Rearranging Eq.~(\ref{eq:n_eh}) expresses the dimensionless functions $\lambda_{\text{e}}\FP{(\xi)}$ and $\lambda_{\text{h}}\FP{(\xi)}$ in Eq.~(\ref{eq:initial-distribution}) as}
\begin{equation}
    \begin{split}
       \lambda_{\text{e}}{(\xi)} &= \frac{n_\text{e}  \alpha}{\int_{\varepsilon_{\text{F}}}^{\varepsilon_\text{max}} dE{\sum\limits_{n\mathbf{k}}}\delta(E-\varepsilon_{n\textbf{k}})\exp\left(-\frac{[\varepsilon_{n\mathbf{k}}-\xi-\Delta/2]^{2}}{2\sigma^{2}}\right)}, \\
        \lambda_{\text{h}}{(\xi)} &= \frac{n_\text{h} \alpha}{\int_{-\varepsilon_\text{max}}^{\varepsilon_{\text{F}}}dE{\sum\limits_{n\mathbf{k}}}\delta(E-\varepsilon_{n\textbf{k}})\exp\left(-\frac{[\varepsilon_{n\mathbf{k}}+\xi+\Delta/2]^{2}}{2\sigma^{2}}\right)}.
    \end{split}\label{eq:lambda_eh}
\end{equation}
Finally, we determine the thermalization time $\tau_{\text{th}}$ through the relation
\begin{equation}
    \frac{P_\text{exc}(\xi,\tau_{\mathrm{th}},T)}{P_\text{exc}(\xi,0,T)}=\frac{1}{e}.
    \label{eq:tau_th}
\end{equation}
{This equation expresses $\tau_{\text{th}}$ as a function of $\xi$ and $T$.}

Let us note that for the materials studied in Refs.~\cite{Yadav2020,Yadav2021} with gap sizes larger than 1 eV and for temperatures below 1000~K, the thermal equilibrium population can be ignored, and $P_\text{exc}(E,t,T)\approx P(E,t,T)$. Since we study here graphene and related few-layer crystals with vanishing or low gap sizes on the order of 100 meV, the thermal contribution needs to be taken into account, and we thus use the excited population $P_\text{exc}(E,t,T)$ in Eq.~(\ref{eq:tau_th}) as well as in the following discussions.

\FP{Regarding our method, the RTA is valid for low excitation intensities, leading to low photoexcited carrier densities $n_\text{e}$ and $n_\text{h}$. It describes the photoexcited carrier relaxation directly to the thermalized state, omitting any intermediate scattering events. In our formalism, temperatures appear in electronic and phononic occupations. We keep electron and phonon baths at the same temperature $T$ at all times, see for instance Eq.~(\ref{eq:self-energy}). The approach thus corresponds to a single-temperature model.}

\section{Results and Discussion}
\label{sec:results}
\begin{table}[h]
\caption{Optimized cell parameters ($a=b$ and $c$), interlayer spacing ($d$) as well as C-C and B-N bond lengths for the studied systems. }
\begin{tabular*}{1.0\columnwidth}{p{0.40\columnwidth}P{0.10\columnwidth}P{0.10\columnwidth}P{0.10\columnwidth}P{0.10\columnwidth}P{0.10\columnwidth}}
	\hline
	\multirow{2}{*}{System} & \multicolumn{5}{c}{{Distance} ({\AA})} \\
	{} & {$ a$} & {{$c$}} & {$d$} & {C-C} & { B-N}\\
	\hline
    (a) $h$-BN/graphene & 2.47 & 20 & 3.21 & 1.43 & 1.43\\
	(b) Monolayer graphene & 2.45 & 16 & --- & 1.41 & ---\\
	(c) Bilayer graphene & 2.45 & 20 & 3.31 & 1.41 & ---\\
	(d) Graphite & 2.45 & 6.61 & 3.31 & 1.41 & ---\\
	\hline
\end{tabular*}
	\label{table:cell_parameters}
\end{table}

Optimized {geometrical} parameters are shown in Table~\ref{table:cell_parameters}. Note that we assume Bernal stacking for \FP{$h$-BN/graphene, bilayer graphene and graphite. The calculated lattice constants for graphite, monolayer and bilayer graphene  are in good agreement with previous {\em ab initio} and experimental values~\cite{CastoNeto:RevModPhy2009,Kong2009,Lebegue2010,Ramasubramaniam2011,Liu_interlayer_2012,gr_ph_layer,razado-colambo_structural_2018}. The difference between the lattice constants of graphene and $h$-BN is less than 3\% \cite{Amorim_2013}, and hence we assume the same hexagonal unit cell for the coupled $h$-BN/graphene system. The optimized lattice parameters for $h$-BN/graphene match well with literature \cite{Giovannetti:PRB2007,Ramasubramaniam2011,gr_hbn_gap}.} 

\begin{figure}[!tb] \centering{}\includegraphics[width=1.0\linewidth]{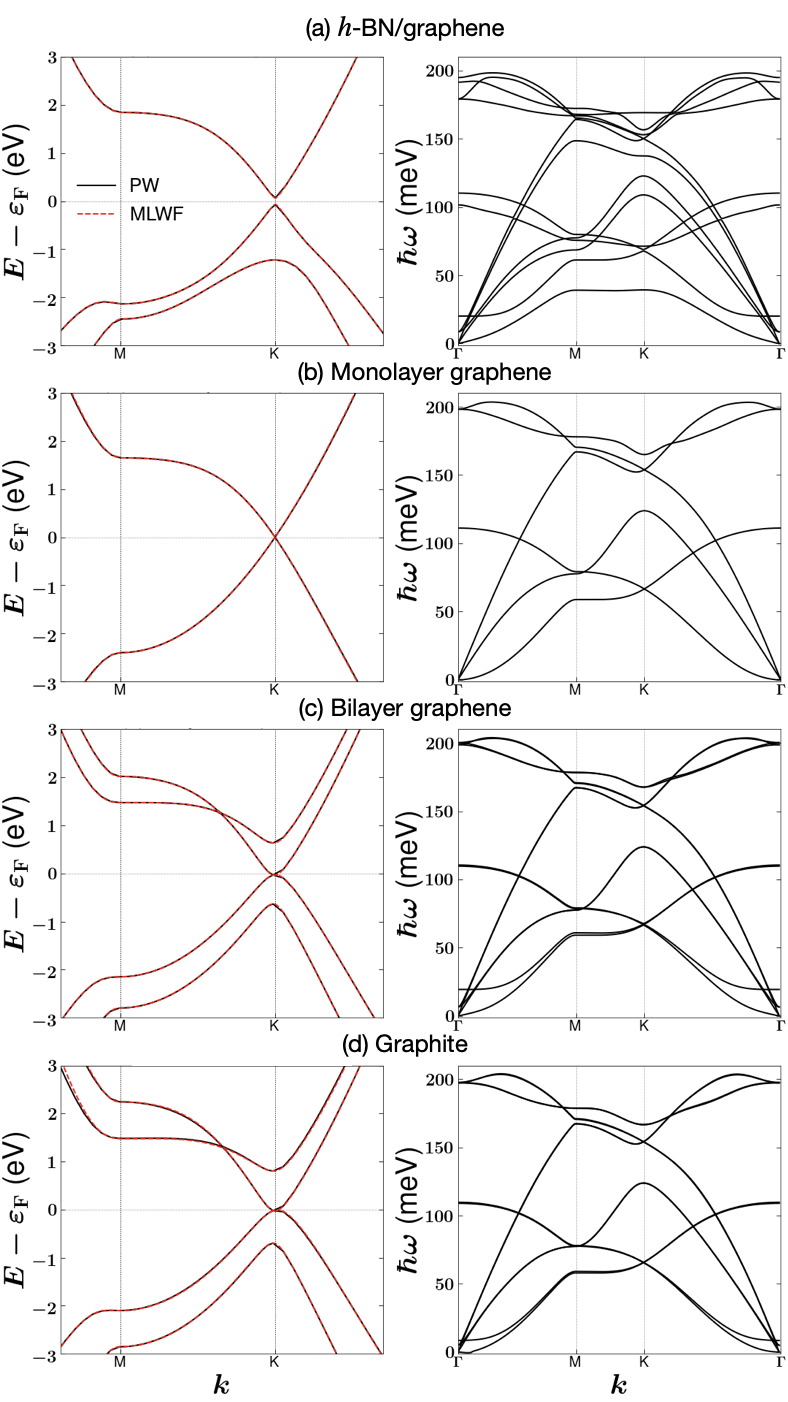} 
	\caption{Electronic (left column) and phononic (right column) band  structures for (a) $h$-BN/graphene, (b) monolayer graphene, (c) bilayer graphene, and (d) graphite along $\mathrm{\Gamma-M-K-\Gamma}$ directions in the irreducible wedge of the BZ.  The electronic band structures are shown for PWs (solid black lines) and MLWFs (red dotted lines) in each case. For $h$-BN/graphene we find a small band gap of around $150$ meV.  Note that we show magnified views of electronic dispersions, and $\Gamma$ points are thus not visible.} \label{fig:bands}
\end{figure}

In Fig.~\ref{fig:bands} we display the electronic and phononic band structures of $h$-BN/graphene, monolayer graphene, bilayer graphene, and graphite. The good agreement of electronic band structures obtained from DFT with PWs and those interpolated using MLWFs demonstrates the quality of the MLWF construction. The electronic band structure of monolayer graphene features the well-known linear energy dispersion around the Dirac point, i.e., around the K point in the  BZ. For hypothetical bilayer graphene without interlayer coupling, {the band structure would be doubled.} Due to interlayer hybridization {the} bands split up. Two pairs of parabolic valence and conduction bands emerge, one pair being energetically well-separated, while the other pair forms {two} bands, which touch each other at the K point and at the Fermi energy~\cite{McCann:PRL2006,CastoNeto:RevModPhy2009}. When graphene is placed on a single $h$-BN layer, the electronic band structure in the vicinity of the Fermi energy is similar to those of graphene. Since $h$-BN is a insulator, its valence and conduction bands are energetically well separated from the linear band crossing at the Dirac point. However, a small bandgap opens up at the Dirac point~\cite{Kan2012,gr_hbn_gap}, because the weak interlayer coupling to $h$-BN distinguishes the carbon atoms in the graphene lattice. \FP{This induces an effective electronic mass term, which opens the gap \cite{Semenoff1984,Kan2012}. Based on our DFT calculations, the electronic band gap opening amounts to around $150$~meV, similar to literature values~\cite{Jung2015,Giovannetti:PRB2007}.} The low-energy band structure of graphite resembles those of bilayer graphene. Analogously, phononic band structures of bilayer graphene and graphite are similar to each other, and their primitive cells contain the same number and type of atoms. In comparison to monolayer graphene with 2 atoms in the unit cell, leading to 3 acoustical (\FP{one of them turning out to be a flexural mode due to the 2D character}) and 3 optical modes, we expect 6 further optical modes in these two materials. In particular, we observe three additional low-energy optical modes at $\Gamma$ below 25~meV, owing to the weak interlayer coupling and a corresponding splitting of what would be acoustical modes in fully decoupled monolayers~\cite{gr_ph_layer}. In the case of the vdW heterostructure $h$-BN/graphene \FP{band degeneracies for acoustical and optical modes are typically removed due to the difference in masses of the constituting atoms B, N and C, in contrast to bilayer graphene and graphite~\cite{Slotman:AnnPhys2014}.}

\begin{figure}[!]
	\includegraphics[width=\linewidth]{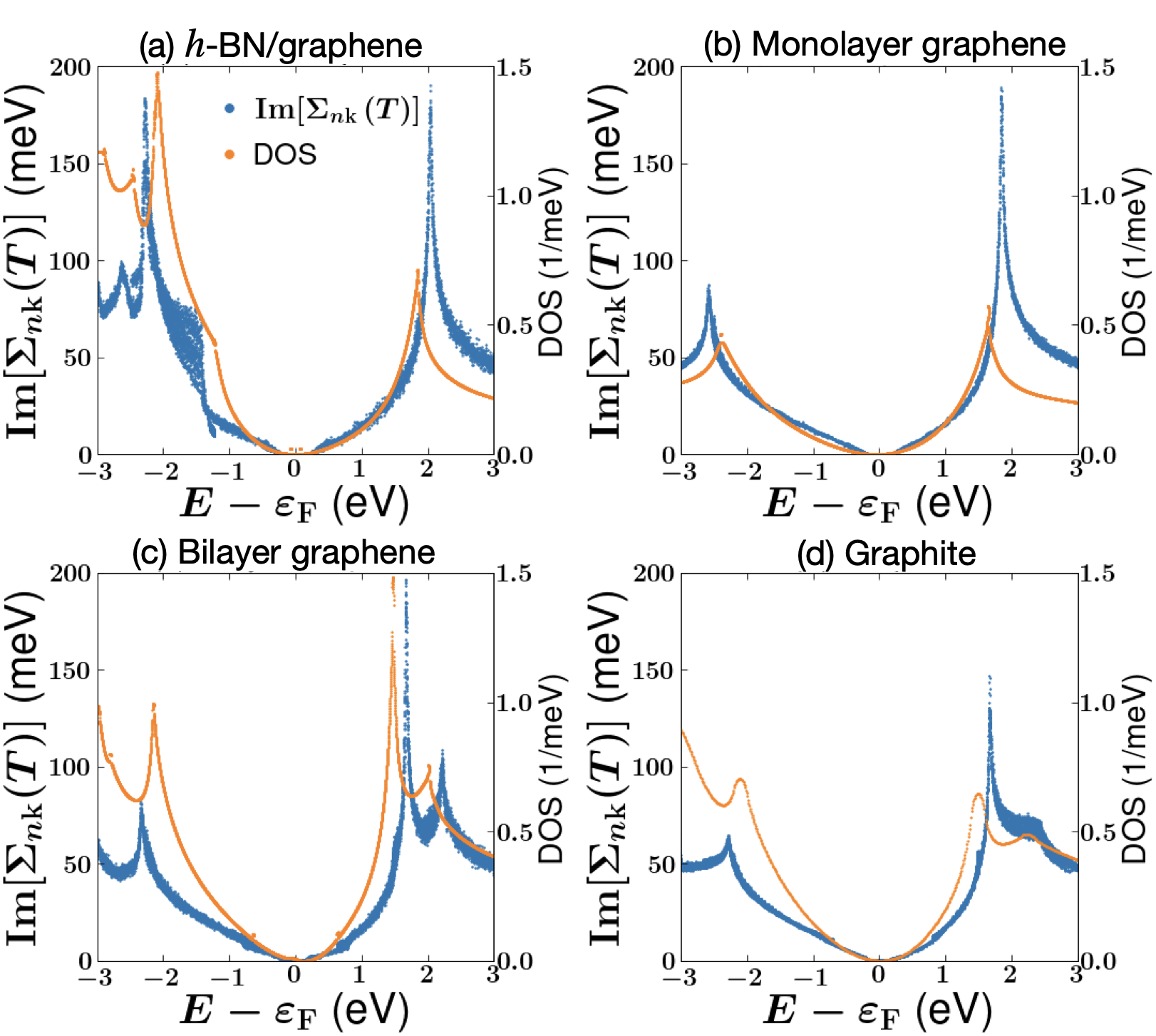} 
	\caption{Imaginary part of the EP self-energy Im$[\Sigma_{n\mathbf{k}}(T)]$ as a function of energy for (a) $h$-BN/graphene, (b) monolayer graphene, (c) bilayer graphene,  and (d) graphite at $T=0$~K, and the corresponding electronic DOS.}
	\label{fig:epc_dos} 
\end{figure}

Having discussed electronic and phononic properties, we analyze $\text{Im}[\Sigma_{n\mathbf{k}}(T)]$ from Eq.~(\ref{eq:self-energy}). It is shown in Fig.~\ref{fig:epc_dos} at $T=0$~K within an energy interval of $\pm 3$~eV around the Fermi energy and compared to the density of states (DOS) for each of the systems under study. We observe that the imaginary part of the self-energy follows the electronic DOS, as the latter essentially represents the phase space for EP scattering. The rapid increase of $\text{Im}[\Sigma_{n\mathbf{k}}(T)]$ with electronic energy away from valence and conduction band edges indicates an efficient energy transfer to phonons for highly excited photocarriers. Electrons or holes with a high energy can scatter repeatedly by phonons and quickly relax to lower energy states, until they reach the band edges. Prominent peaks around $\pm2$~eV in Fig.~\ref{fig:epc_dos} can be attributed to the slowly varying energy bands around the M point in the BZ. 

Similar to our previous study on graphene~\cite{Yadav2019}, we find that photocarriers with an energy below the optical phonon energies of about 150~meV relax at a much slower rate than above. In order to focus on this interesting region, we will therefore perform our calculations for the remainder of this work within the 300~meV energy window of conduction and valence band edges, i.e.\, within about $\pm 300$~meV around $\varepsilon_\text{F}$.

\begin{figure}[!]
	\includegraphics[width=\linewidth]{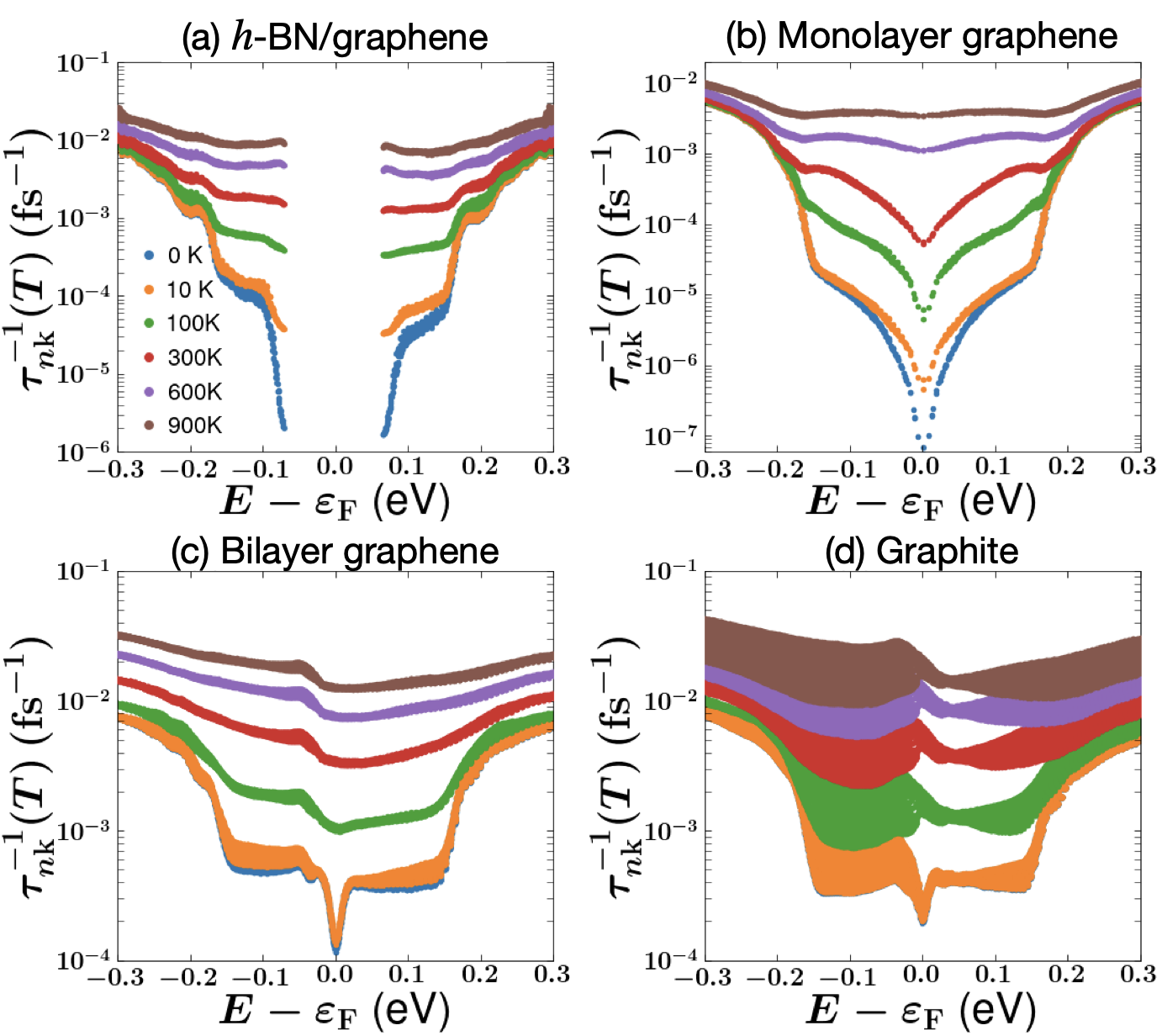} 
	\caption{EP scattering rates $\tau_{n\mathbf{k}}^{-1}(T)$ as a function of energy for (a) $h$-BN/graphene, (b) monolayer graphene, (c) bilayer graphene,  and (d) graphite,  evaluated at different temperatures ranging from $T=0$ to $900$~K.}
	\label{fig:epc_scattering} 
\end{figure}

The EP scattering rates $\tau^{-1}_{n\mathbf{k}}(T)$ of Eq.~(\ref{eq:tau-nk}) are shown in Fig.~\ref{fig:epc_scattering} as a function of energy for different temperatures in the range of $0$ to $900$~K. These temperatures correspond to a realistic range, given that melting points for graphene and $h$-BN have been reported to be as large as 4900~K~\cite{melting_graphene} and 3300~K~\cite{melting_hbn}, respectively. For all the analyzed materials, we observe that $\tau^{-1}_{n\mathbf{k}}(T)$ varies by orders of magnitude with regard to energy and temperature. 

Let us first concentrate on the low-temperature limit of $T = 10$~K. For graphene, scattering rates increase strongly as a function of energy around $\varepsilon_\text{F}=0$, highlighting inefficient EP scattering until the highest phonon branches at $150-200$~meV are reached. Above around 150~meV the scattering rates increase abruptly due to emission of optical phonons~\cite{Winnerl2011, Yadav2019}. For the multilayer structures we see a more complex behavior of $\tau^{-1}_{n\mathbf{k}}(T)$ with  step-like features within the energy interval of $\pm 150$~meV around the Fermi energy. \FP{For bilayer graphene and graphite there exist pronounced plateaus in energy from 20 to 150~meV, while for $h$-BN/graphene we see two plateaus, which range in absolute energies from 95 to 150~meV and 175 to 200~meV. Taking into account the offset $\Delta/2=75$~meV due to the band gap in $h$-BN/graphene, we} attribute this behavior \FP{in bilayer graphene, graphite and $h$-BN/graphene} to the low-lying optical phonons with energies as low as 5~meV (see Fig.~\ref{fig:bands} and Table~\ref{table:k2g_scattering}).

For all the materials studied, EP scattering rates furthermore exhibit a strong temperature dependence below 150~meV and a weaker one above, the energy threshold coinciding again with the high-lying optical phonon energies. In our single-temperature model, where electrons and phonons are described by the same $T$, a higher temperature leads to a larger thermal smearing of both electronic and phononic bath occupations. The scattering rates, hence, increase monotonically with temperature, as illustrated in Fig.~\ref{fig:epc_scattering}, where the rates are plotted for $T=0, 10, 100, 300, 600$ and $900$~K. As discussed exemplarily for $10$~K before, scattering rates for temperatures of 300~K or below show a pronounced energy dependence and promptly rise, as we move away from the Fermi level or the band edges. At $T\leq300$~K and energies below 150~meV all systems feature rather low scattering rates. Elevated temperatures increasingly wash out the energy dependence of $\tau^{-1}_{n\mathbf{k}}(T)$, see especially the cases of $T=600$ and $900$~K, which we attribute to phonon absorption. For monolayer graphene, shown in Fig.~\ref{fig:epc_scattering}(b), the scattering rate near the Fermi level increases from below $10^{-6}$~fs$^{-1}$ at $T=10$~K by some three orders of magnitude to $10^{-3}$~fs$^{-1}$ at $T=600$~K. The $h$-BN/graphene vdW heterostructure behaves in a similar manner at the valence and conduction band edges. The corresponding values for bilayer graphene and graphite are increased to a lesser extent from $10^{-3}$~fs$^{-1}$ at $T=10$~K by around one order of magnitude to $10^{-2}$~fs$^{-1}$ at $T=600$~K. We explain this difference by the parabolic dispersion of low-energy electron bands in multilayer graphene systems \cite{McCann:PRL2006,CastoNeto:RevModPhy2009}, which lifts the mismatch between electronic and acoustical phonon velocities that exists in monolayer graphene.  The scattering rates for graphite feature a comparatively large spread. Since we observe the fuzziness only in graphite, we attribute it to the lack of 2D confinement. Let us also note that the scattering of holes and electrons is very much comparable in monolayer graphene, in accordance with the highly symmetric DOS visible in Fig.~\ref{fig:epc_dos}. In the case of $h$-BN/graphene, bilayer graphene, and graphite scattering rates are slightly higher for holes than for electrons, a feature that is also partly reflected in the DOS plots of Fig.~\ref{fig:epc_dos}.

\begin{figure}[!t]
	\includegraphics[width=\linewidth]{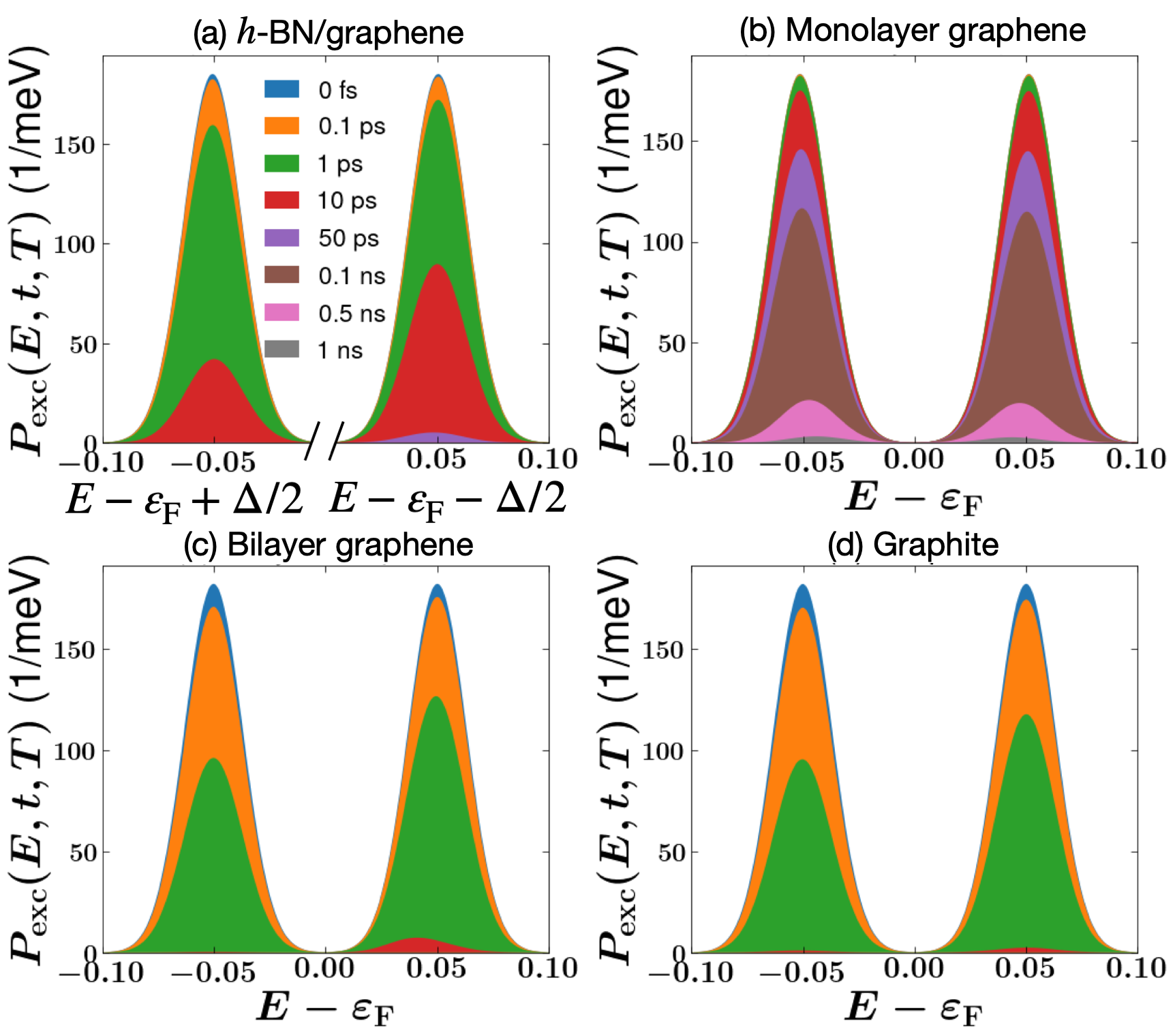} 
	\caption{Time-dependent \FP{excited} photocarrier population $P_{\rm exc}(E, t, T)$ as a function of energy, measured in eV, for electrons and holes at an excess energy $\xi=50$~meV and temperature $T=10$~K for (a) $h$-BN/graphene, (b) monolayer graphene, (c) bilayer graphene,  and (d) graphite.}
	\label{fig:population}
\end{figure}

The time-dependent population $P_\text{exc}(E, t, T)$ of excited carriers is plotted in Fig.~\ref{fig:population} at $T=10$~K for an excess energy $\xi=50$~meV, i.e.\ electron energy of 50~meV above the conduction band minimum (CBM) and hole energy {of 50~meV} below the valence band minimum (VBM). {At $t=0$, we} excite $10^{13}$~cm\textsuperscript{-2} carriers in the 2D monolayer and bilayer systems and ${1.5 \times}10^{13}$~cm\textsuperscript{-3} in graphite {[see Eqs.~(\ref{eq:n_eh}) and (\ref{eq:lambda_eh})]}. While monolayer and $h$-BN/graphene show the slowest thermalization dynamics, bilayer graphene and bulk graphite thermalize by orders of magnitude faster. \FP{Initial photoexcited carrier populations of electrons and holes are identical in all materials due to the excitation assumed [see Eq.~(\ref{eq:initial-distribution})]. At later times, we find holes to relax faster than electrons for $h$-BN/graphene, bilayer graphene and graphite in Fig.~\ref{fig:population}(a,c,d). This is clearly visible from the different peak heights that develop with time. Only for graphene electrons and holes relax at a similar pace. This behavior is expected from the scattering rates, displayed in Fig.~\ref{fig:epc_scattering}. There is an additional asymmetry, occurring in electron and hole populations separately in the course of time, due to the energy dependence of the scattering rates, see for instance the slight shift of peak maxima towards $\varepsilon_\text{F}$ in Fig.~\ref{fig:population}.}

\begin{figure}[!t]
	\centering{}\includegraphics[width=1.0\linewidth]{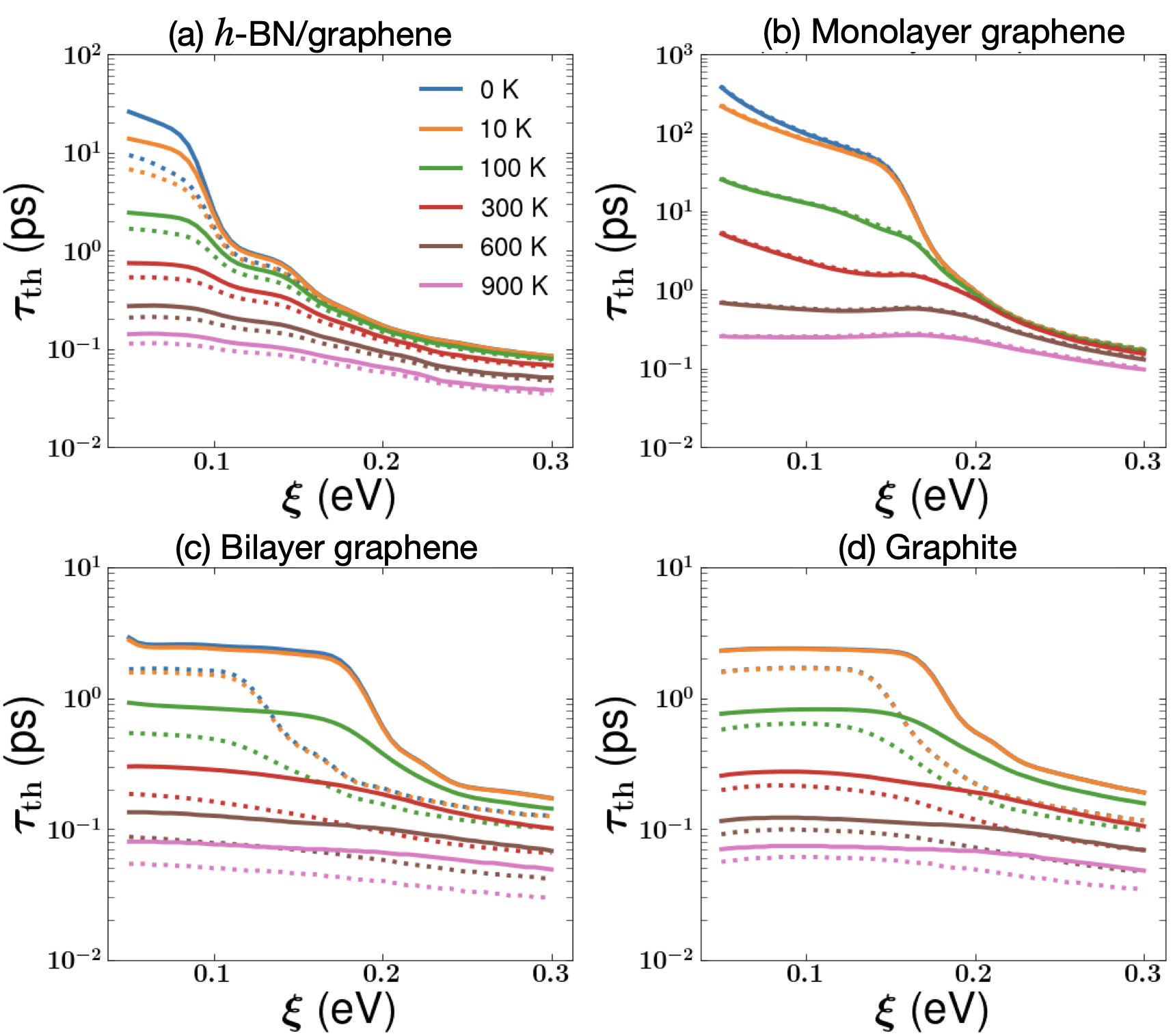}
	\caption{Thermalization times $\tau_{\mathrm{th}}$ of electrons (solid lines) and holes (dashed lines) as a function of excess energy $\xi$ at different temperatures for (a) $h$-BN/graphene, (b) monolayer graphene, (c) bilayer graphene,  and (d) graphite.}
	\label{fig:t-th} 
\end{figure}

In order to obtain an overview of photocarrier dynamics, we summarize the thermalization times $\tau_{\text{th}}$ [see Eq.~(\ref{eq:tau_th})] of electrons and holes as a function of excess energy at different $T$ in Fig.~\ref{fig:t-th}. For excitations with $\xi<150$~meV, i.e.\ below the high-lying optical phonon energies of monolayer graphene, the thermalization time decreases rapidly with increasing temperature in all materials studied. Simultaneously, the thermalization time is substantially {extended} for the systems containing only a single graphene layer as compared to the multilayer graphene stacks, if temperatures are sufficiently low. Finally, beyond the highest optical phonon energy of around 200~meV, carriers thermalize quickly in all materials, demonstrating only a weak dependence on temperature. Thermalization times then drop to a few hundred femtoseconds or less, sharply contrasting those at smaller excess energies {$\xi$}. 

In Fig.~\ref{fig:t-th} graphene shows nearly identical thermalization times for electrons and holes (with dashed and solid lines coinciding). For the other systems, electrons thermalize more slowly than holes, as expected from Figs.~\ref{fig:epc_scattering} and \ref{fig:population}. The overall dependence of $\tau_\text{th}$ on $\xi$ is very similar for all carbon-based systems, i.e.\ for graphene, bilayer graphene and graphite, and exhibits a plateau region followed by a sharp drop starting from around 150~meV for $T<300$~K. \FP{The plateau is somewhat inclined for graphene but much flatter for bilayer graphene and graphite.}  The magnitude of thermalization times at low temperature and $\xi$ is reduced by two orders of magnitude in bilayer graphene and graphite as compared to monolayer graphene. Maximum values of thermalization times for the $h$-BN/graphene heterostructure at low $T$ are reduced by around an order of magnitude as compared to those of monolayer graphene. Apart from this quantitative difference, we observe a complex dependence on the excitation energy, where $\tau_\text{th}$ decreases in two steps at around 100 and 150~meV, particularly well visible at $T<300$~K and similar to the plateau features found for the EP scattering rates in Fig.~\ref{fig:epc_scattering}. \FP{As before, the deviations from the behavior of monolayer graphene in the multilayer systems arise from electron and phonon band hybridization, the resulting emergence of new low-energy optical phonon modes and the modified EP interactions.}

\begin{figure}[!t]
    \centering
    \includegraphics[width=1\columnwidth]{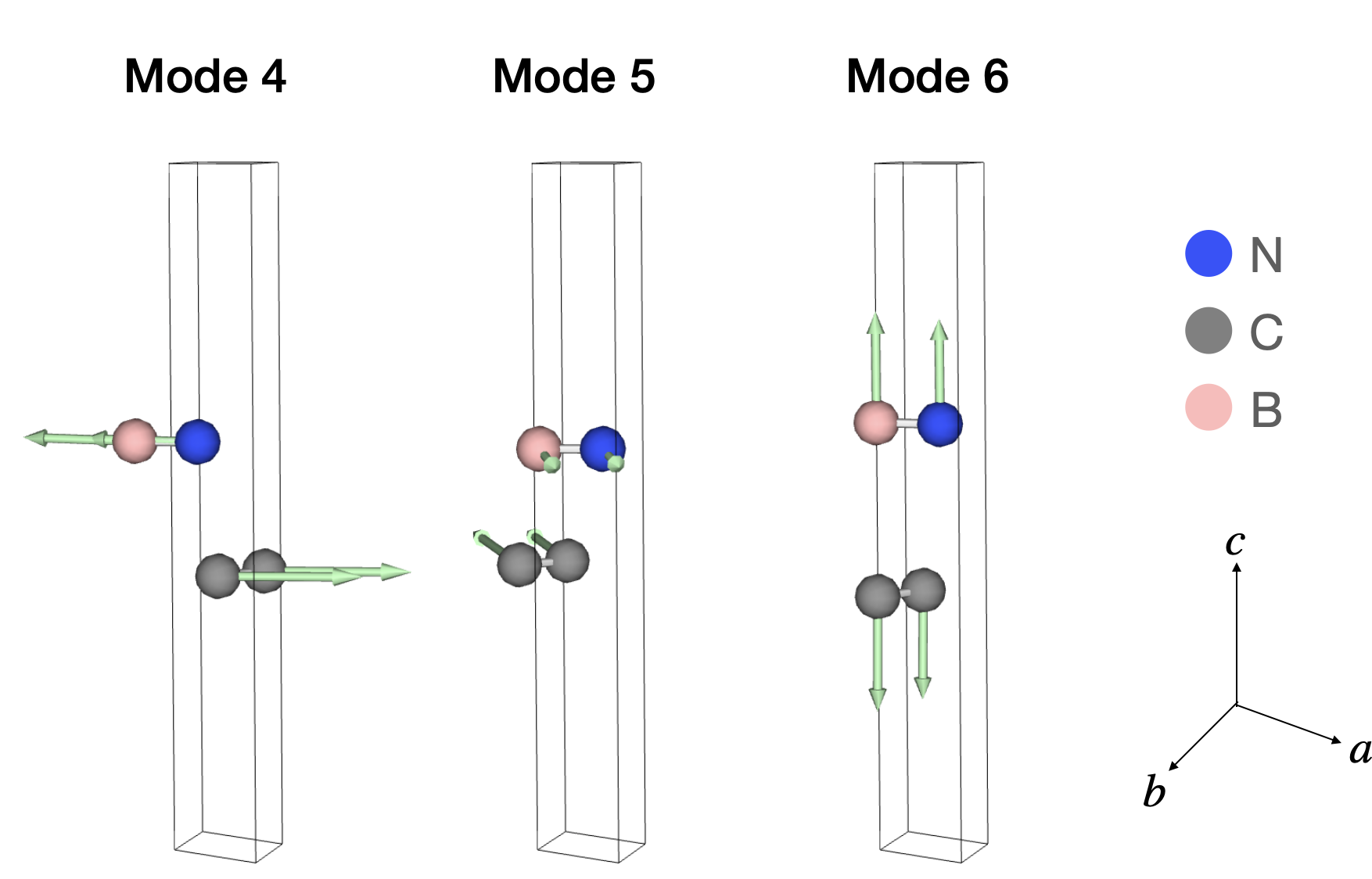}
    \caption{Optical phonon modes with energies below 50~meV for the $h$-BN/graphene system at the $\Gamma$ point. \FP{The color code of atoms and the coordinate system are shown on the right side}.}
    \label{fig:normal_modes16}
\end{figure}

In the previous paragraphs, we have seen that the absence of low-energy optical modes leads to the thermalization bottleneck in graphene. Such low-energy optical modes are however present in the other bi- and multilayer systems studied here. {In order to understand the role of optical phonon modes with energies below 50~meV in the thermalization, we have computed the corresponding EP scattering rates for electrons at the Dirac point (K) due to the phonons at $\Gamma$ for a temperature of 10~K. In this way, we describe the most relevant low-energy processes, i.e.\ the interaction of electrons  at the K valley with phonons with vanishing momentum. Let us emphasize that due to the crossing of phonon bands (see Fig.~\ref{fig:bands}), a fully mode-resolved description of scattering rates on the whole BZ is not possible, hence we resort to a single $\mathbf{q}$ point. In Table~\ref{table:k2g_scattering} we present the EP scattering rates for all of the materials and all those optical modes at $\Gamma$ with energies below 50~meV.} Note that the first three acoustical phonons are omitted, since they yield no contribution. The reason is that EP couplings vanish for pure translations. {In Fig.~\ref{fig:normal_modes16}, we furthermore display the form of the three energetically lowest optical phonon modes at the $\Gamma$ point for $h$-BN/graphene, i.e., excluding  the acoustical modes. The modes for the other multilayer systems bilayer graphene and graphite are qualitatively similar, in particular modes 4 and 5 are always of in-plane character while mode~6 describes an out-of-plane motion.}

Table~\ref{table:k2g_scattering} reveals that the optical phonon modes at $\Gamma$ with energies below 50~meV contribute to the scattering of electrons at K for $h$-BN/graphene, bilayer graphene, and bulk graphite. Their contributions cause a faster relaxation of hot carriers in contrast to monolayer graphene, where such low-energy modes are missing. For bilayer graphene and graphite, the low-energy in-plane optical modes ({modes 4 and 5}) are mostly responsible for the scattering, as indicated by the values of $\tau_{1\text{K}}^{-1}$ in Table~\ref{table:k2g_scattering}. Rates of in-plane modes are comparable for these two materials, and thermalization in bilayer graphene and graphite therefore takes place on similar time scales in agreement with Fig.~\ref{fig:t-th}. While for graphite the contribution of the out-of-plane optical mode is very small, it is much higher for bilayer graphene and further enhanced for the $h$-BN/graphene vdW heterostructure. This shows that the EP scattering due to interlayer coupling is more pronounced in 2D systems relative to their bulk counterparts. \FP{Note also that the energy of the out-of-plane mode 6 is substantially reduced in graphite as compared to $h$-BN/graphene and bilayer graphene. A similar softening is also observable for the in-plane modes, although their energies remain more similar throughout the considered multilayer materials.} {Remarkably, for $h$-BN/graphene the contribution of the out-of-plane mode to the EP scattering is comparable to the in-plane ones, but the rates $\tau_{1\text{K}}^{-1}$ for modes 4-6 are overall around two orders of magnitude lower than those of the important in-plane modes for bilayer graphene and graphite. This explains why the thermalization times in $h$-BN/graphene are longer than in these two carbon-based materials and why the thermalization bottleneck in $h$-BN/graphene is largely preserved.}

\MT{Let us now relate our outcomes to experimental results on graphene and graphene-based heterostructures. Dawlaty \emph{et al.}~\cite{Dawlaty2008Jan} have measured the carrier relaxation times in epitaxial graphene layers on SiC wafers using ultrafast optical pump-probe spectroscopy and observed two distinct time scales associated with the relaxation of nonequilibrium photogenerated carriers: an initial fast relaxation transient in the 70 to 120~fs range, followed by a slower relaxation process in the $0.4$ to $1.7$~ps range. These fast and slow time constants are related to carrier-carrier and carrier-phonon scattering processes in graphene. 
Note that epitaxial "graphene" typically does not represent a single layer but is \FP{in fact a stack of a few graphene layers}.
Our model predicts that the phonon-related thermalization time does not strongly depend on temperature in the optical excitation limit ($\xi>0.5$~eV \FP{in Ref.~\cite{Dawlaty2008Jan}}) and is indeed around 0.1 ps for bilayer graphene and graphite.
In contrast, the carrier relaxation time in $h$-BN-encapsulated graphene has been measured for near infrared excitations ($\xi\approx 0.4$~\FP{eV}) at temperatures
in the range from 10 to 300~K \cite{Jadidi2016Dec} and was found to be one order of magnitude higher, as expected from our Fig.~\ref{fig:t-th}.

Optically excited electrons, injected from graphene into an adjacent 2D semiconductor, have been studied in Ref.~\cite{Chen2019} at temperatures ranging from 100 to 300~K. The electronic gap of the semiconductor filters out low-energy (i.e.\ thermalized) electrons so that only non-thermalized electrons are collected, see also Ref.~\cite{Yadav2020}.  
The measurement shows 25~fs electron injection time from graphene to the 2D semiconductor with up to 50\% quantum yield. Our model predicts a thermalization time of about 100~fs at the relevant excitation energies, which matches well with the expectations. 
Note that the measured quantum yield shows strong dependence on photon energy but remains nearly constant with varying photon density (fluence).
This is in good agreement with our model, \FP{predicting a thermalization time that strongly depends on $\xi$.} At the same time, it independently justifies our assumption of linear response, neglecting \FP{possible complex nonlinear} intensity dependences.}

\FP{Finally, let us discuss different possibilities to improve the theoretical approach presented here. We note that DFT methods using exchange correlation functionals based on LDA do not fully capture the Kohn anomalies in graphene and graphite \cite{Lazzeri2008Aug,Tong2021}. Kohn anomalies have been argued to be responsible for a prevalent contribution of the high-energy optical phonons to electronic thermalization in graphite \cite{Piscanec2004,Tong2021}. Despite the shortcomings, our LDA calculations reproduce this crucial relaxation mechanism. Nevertheless, an improved description of Kohn anomalies through $GW$ corrections appears to be worthwhile in future work \cite{Lazzeri2008Aug,Tong2021}. Furthermore, the use of polar corrections~\cite{Zhou2016Nov,Zhou2018Nov} might be advised especially for the $h$-BN/graphene system. We checked that they do not lead to significant changes in the EP scattering rates of Fig.~\ref{fig:epc_scattering}. We did not consider polar corrections here, because for $h$-BN/graphene they resulted in two imaginary frequencies in the phonon dispersion at the $\Gamma$ point. Finally, we employ a single temperature for electrons and phonons and use the RTA. Instead, coupled electron and phonon distributions may be propagated in time, involving a reevaluation of scattering rates in every time step \cite{Tong2021}. In this way, the occupation of all electronic and phononic modes in the BZ is tracked, representing a "multi-temperature" model. In addition, methods may be utilized to better describe the quantum behavior of electrons and phonons \cite{kadanoff2014entropy, KB_equation}.} 

\vspace{1 cm}
\begin{table*}[!bt]
\caption{Phonon modes $p$ at $\mathbf{q}=\Gamma=0$ with energy $\hbar \omega_{p\Gamma} < 50$~meV,
and the scattering rates $\tau_{n\mathbf{k}}^{-1}(T)$ at $T=10$~K for electrons in band $n=1$ at the Dirac point $\mathbf{k}=$K due to the interaction with the corresponding phonon mode $p$ at $\mathbf{q}=\Gamma$. No data is shown for monolayer graphene, since all of its optical phonon modes are located at energies above 50~meV. In order to provide scattering rates that are consistent with the data shown in Fig.~\ref{fig:epc_scattering}, we have {divided} the rates obtained by \textsc{perturbo} for $\mathbf{k}=\text{K}$ and $\mathbf{q}=\Gamma$ 
by the factor of $10^7$, which is the number of total $\mathbf{q}$ points used previously.
}
\begin{tabular*}{1.0\linewidth}{p{0.10\linewidth}P{0.1\linewidth}P{0.1\linewidth}P{0.1\linewidth}P{0.1\linewidth}P{0.1\linewidth}P{0.1\linewidth}P{0.1\linewidth}P{0.1\linewidth}}
\hline
	\multirow{1}{*}{\bf Mode $p$} & 
	\multicolumn{2}{c}{\bf $h$-BN/graphene } &
	\multicolumn{2}{c}{\bf Monolayer graphene  } &
	\multicolumn{2}{c}{\bf Bilayer graphene} &
	\multicolumn{2}{c}{\bf Graphite} \\
	{} & {$\hbar \omega_{p\Gamma}$ (meV)} & {$\tau_{1\text{K}}^{-1}$ (fs$^{-1}$)} & {$\hbar \omega_{p\Gamma}$ (meV)}  & {$\tau_{1\text{K}}^{-1}$ (fs$^{-1}$)} & {$\hbar \omega_{p\Gamma}$ (meV)}  & {$\tau_{1\text{K}}^{-1}$ (fs$^{-1}$)} & {$\hbar \omega_{p\Gamma}$ (meV)}  & {$\tau_{1\text{K}}^{-1}$ (fs$^{-1}$)} \\
\hline \\
4    & {8.57} & { $2.20 \times 10^{-9}$} & --- & --- & $6.48$ & $2.64 \times 10^{-7}$  & $4.93$ & $5.80 \times 10^{-7}$   \\
5    & {8.57} & { $3.21 \times 10^{-9}$} & --- & --- & $6.48$ & $2.60 \times 10^{-7}$  & $4.93$ & $5.78 \times 10^{-7}$   \\
6    & 20.07 & {$4.49 \times 10^{-9}$} & --- & --- & $19.3$ & $3.36 \times 10^{-11}$ & $8.48$ & $4.59 \times 10^{-16}$  \\
\hline 
\end{tabular*}
	\label{table:k2g_scattering}
\end{table*}

\section{Conclusions}\label{sec:conclusion}
In conclusion, we carried out a detailed study of EP-induced scattering rates and the related photocarrier thermalization in the $h$-BN/graphene vdW heterostructure. In this context, monolayer and bilayer graphene as well as bulk graphite served as reference systems. Combining DFT and DFPT with a MLWF interpolation, we calculated the excited carrier relaxation times from first principles, i.e.\ without free parameters, for a large range of temperatures and excess energies.

We find that EP scattering rates differ significantly for the investigated systems at low temperatures and at excitations below around $150$~meV. At temperatures above 600~K and at higher excitation energies, the scattering rates turn out to be quite comparable.

Monolayer graphene exhibits an extremely low scattering rate at low temperatures and at low excitation energies, resulting from its linear band dispersion and the absence of low-energy optical phonon modes, which leads to a slow cooling of excited carriers. In contrast to bilayer graphene, where the photo-carriers thermalize similarly rapidly as in graphite, replacing one graphene layer with $h$-BN restores the hot-carrier thermalization bottleneck found in monolayer graphene. We attribute this to a comparatively weak coupling of low-energy optical phonon modes to the electronic states in this particular heterostructure. Suppressed EP coupling along with a low impurity concentration at the $h$-BN/graphene interface may be responsible for the large electron mobility, typically observed in such structures. Understanding interfacial coupling mechanisms, we foresee deliberate approaches to design functional electronic devices based on vdW heterostructures.

\section*{Acknowledgments}
The authors thank Goki Eda for stimulating discussions. The computational facilities provided by the Okinawa Institute of Science and Technology (OIST) Graduate University are acknowledged. This research is also supported by the Ministry of Education, 
Singapore, under its Research Centre of Excellence award to the Institute for Functional Intelligent Materials (I-FIM, project No. EDUNC-33-18-279-V12).

%


\end{document}